\documentclass[10pt,journal,compsoc]{IEEEtran}
\usepackage{multirow}
\usepackage{amssymb}
\usepackage{booktabs}
\usepackage{threeparttable}
\usepackage{enumitem}
\usepackage[framemethod=tikz]{mdframed}
\usepackage{tikz}
\usepackage{tipa}
\usepackage{xtab}
\usepackage{mfirstuc}
\usepackage{todonotes}
\usepackage{soul}
\usepackage{colortbl}
\usepackage{xspace}
\usepackage{xcolor}

\presetkeys%
    {todonotes}%
    {inline}{}
\usepackage{url}
\usepackage{relsize}
\def\mycolor{blue}
\def\histogram#1{
  \begin{tikzpicture}[xscale=0.04, yscale=0.04]
    \useasboundingbox (-0.2,-0.2) rectangle (16.2,3.2); % The draw option is given just for demonstration.
    \begin{scope}[ycomb, yscale=0.6]
      \draw[fill=\mycolor!100!black!10, \mycolor!100!black!10] (0,0) rectangle (18,10); 
      \draw[xshift=-14pt,  \mycolor!70!black!80, line width=1.15] plot[] file {#1}; 
      \draw[\mycolor!100!black!25, line width=0.2] (-0.1,-0.4) rectangle (18.1,10.4);  
   \end{scope}
\end{tikzpicture}
}

\newcommand{\q}[1]{\emph{``#1''}}

\newcommand{\histno}[2]{({\histogram{histogram/#1.tex}, #2 mentions})\xspace}
\newcommand{\nfr}[1]{\textsc{#1}}
\newcommand{\nfrno}[1]{\textsc{#1}}
\newcommand{\cm}[1]{\nfr{Configurability}}

\newcommand{\hll}[1]{#1}

\newcommand{\hghlt}[1]{#1}

\newcommand{\tied}[1]{T#1}%\tnote{*}}

\usepackage[many]{tcolorbox}
\newtcolorbox{mybox}[1][]{
  breakable,
  title=#1,
  colback=white,
  colbacktitle=white,
  coltitle=black,
  fonttitle=\bfseries,
  bottomrule=0pt,
  toprule=0pt,
  leftrule=3pt,
  rightrule=3pt,
  titlerule=0pt,
  arc=0pt,
  outer arc=0pt,
  colframe=black,
}

\begin{document}

\author{Colin Werner,
        Ze Shi Li,
        Derek Lowlind,
        Omar Elazhary,
        Neil Ernst,
        and Daniela Damian} %<-this % stops a space
\markboth{IEEE Transactions on Software Engineering}%
{Werner \MakeLowercase{\textit{et al.}}: GDPR Journal}

% \title{Continuous Non-Functional Requirements: Practices, Opportunities, and Trade-offs} %for Small, Agile Organizations}% Deal With Non-Functional Requirements}% in Continuous Software Delivery Settings}
\title{Continuously Managing NFRs: Opportunities and Challenges in Practice}

\IEEEtitleabstractindextext{%
\begin{abstract}%175 words
Non-functional requirements (NFR), which include performance, availability, and maintainability, are vitally important to overall software quality. 
However, research has shown NFRs are, in practice, poorly defined and difficult to verify.
%In particular, NFRs are often ignored in agile settings. 
%Olsson and Bosch: "While the ability to continuously deploy new software functionality creates new business opportunities and extends the concept of agile practices, it presents a number of challenges."
%Guzman/Franch: A recent evolutionary step from agile software development is rapid and continuous software engineering, which refers to the organizational capability to develop, release, and learn from software in rapid cycles.
Continuous software engineering practices, which extend agile practices, emphasize fast paced, automated, and rapid release of software that poses additional challenges to handling NFRs.
In this multi-case study we empirically investigated how three organizations, for which NFRs are paramount to their business survival, manage NFRs in their continuous practices.
We describe four practices these companies use to manage NFRs, such as offloading NFRs to cloud providers or the use of metrics and continuous monitoring, both of which enable almost real-time feedback on managing the NFRs. 
However, managing NFRs comes at a cost---as we also identified a number of challenges these organizations face while managing NFRs in their continuous software engineering practices. 
For example, the organizations in our study were able to realize an NFR by strategically and heavily investing in configuration management and infrastructure as code, in order to offload the responsibility of NFRs; however, this offloading implied potential loss of control.
Our discussion and key research implications show the opportunities, trade-offs, and importance of the unique give-and-take relationship between continuous software engineering and NFRs. \\
Research artifacts may be found at \url{https://doi.org/10.5281/zenodo.3376342}.

\end{abstract}

% <concept_id>10011007.10011074.10011075.10011076</concept_id>
% <concept_desc>Software and its engineering~Requirements analysis</concept_desc>
% <concept_id>10011007.10011074.10011081.10011082</concept_id>
% <concept_desc>Software and its engineering~Software development methods</concept_desc>
% <concept_significance>300</concept_significance>
% \ccsdesc[500]{Software and its engineering~Requirements analysis}
% \ccsdesc[300]{Software and its engineering~Software development methods}

}

\maketitle
\IEEEdisplaynontitleabstractindextext

% \fi
\IEEEpeerreviewmaketitle

\section{Introduction}
Non-functional requirements (NFRs), also known as quality attribute requirements, represent attributes or constraints on a system \cite{glinz}.
NFRs are very important in software projects, greatly influencing underlying software architecture \cite{chen2012characterizing}.
For many software organizations, particularly in today's increasingly service-oriented, fast paced software development marketplace, NFRs such as system uptime, code maintainability, and responsiveness, are vital to success \cite{caracciolo2014software}. 
For example, a recent software outage at Amazon was estimated to cost 99 million USD for the 63 minutes Amazon's site was down \cite{cbr-aws}. 
The majority of software organizations \cite{ci_stats} now use continuous software engineering (CSE) \cite{fowler, Fitzgerald2014} which involves rapid and frequent builds, automated tests, as well as transparency of the build and verification process \cite{fowler}. 

Despite the great benefits organizations reap through CSE \cite{hilton2016usage}, NFRs have rarely been explored in the context of CSE \cite{behutiye2020management}. % \cite{yu2020utilising}. 
Studies of practice demonstrate NFRs are inherently difficult to explicitly express \cite{Eckhardt:2016} and even more difficult to verify or validate \cite{Nuseibeh2000}, whether they are part of a formal requirements specification or an agile user story.
In agile settings NFRs present more engineering difficulties than functional requirements \cite{Alsaqaf2019}. 
Continuous settings that focus on automated and rapid release of software compound these challenges, for example in A/B testing \cite{mattos2020}. 
This leads to long-term sustainability issues and mounting technical debt \cite{chen2014architecture} resulting in costly and unnecessary rework \cite{wagner_literature_2006}.

\hll{Fowler's original definition of CSE practices \cite{fowler} only mentions testing of NFRs. 
In addition to testing, however, managing, modeling, and eliciting NFRs, or \emph{quality} requirements, are all central to CSE's overall goal of improving software quality \cite{fitzgerald2017continuous}.}
The software engineering literature lacks empirical evidence on how organizations can verify, let alone manage, implement, and realize an NFR by leveraging CSE \cite{yu2020utilising}.
%\omar{Isn't the previous statement because CSE practices initially focus on mitigating integration issues?}
 
%CSE \cite{fowler, Fitzgerald2014}, rooted in agile methodologies, require an organization to support rapid, frequent builds; automated tests; and transparency of the build and test process \cite{fowler}. 
%As NFRs may be difficult to verify, CSE poses significant challenges to the engineering of NFRs.
%Despite the importance and overarching implications of NFRs there is currently a
%In particular, CSE often leads to short-term functional prioritization over NFRs \cite{gralha_evolution_2018}.

%Research in general has understudied smaller organizations \cite{aranda2007requirements}, although a small organization is no less concerned with NFRs. 
%For example, application uptime may be one of the most important requirements to a small web startup. 
%Yet, the ability to deal with these requirements in the smaller, less resource-rich setting of such an organization is unknown \cite{alsaqaf2017quality}. 
%For example, practitioners, and researchers, are unclear on the best approach in automating NFRs, or which NFRs need to be handled internally versus being offloaded to third parties. \todo{can we say something about which NFRs, of our list, are being offloaded in particular? }
\hll{In this paper we report on an exploratory study that aims to fill this gap through empirical insights on the relationship between NFRs and CSE, from an industrial multiple-case study \cite{yin_case_2002} of managing NFRs in CSE. }
%on the practices and associated challenges in handling NFRs.
Two research questions guided our study:
\begin{enumerate}[label=\textbf{RQ\arabic*},leftmargin=*]
    \item How do CSE organizations manage NFRs? 
    \item What challenges does CSE introduce when managing NFRs?
\end{enumerate}
We investigate the CSE practices at three organizations for which NFRs, such as performance, security and availability, are paramount to their business survival. 
We collected data by observing %\cite{potts_research} 
employees in a number of immersive visits we conducted at these organizations, and through 18 semi-structured follow-up interviews.
%We performed our data analysis using thematic analysis \cite{cruzes_recommended_2011}.
We interacted with a variety of roles that are directly involved with or impacted by how NFRs were handled in CSE, including developers, DevOps engineers, QA testers, and managers.
Following thematic analysis \cite{cruzes_recommended_2011} on our rich qualitative data, we identify a set of practices that these organizations employ to handle their NFRs (RQ1), e.g. letting someone else manage an NFR by offloading it to a third-party.
%\todo{this is not the most interesting one. what is the most interesting practice?} e.g. leveraging the quick feedback loop to create transparent metrics used across the organization.
The answer to RQ2 exposes the challenges CSE introduces when managing NFRs, e.g. the fast pace of development led to a decrease in the shared understanding of NFRs.

The contributions of this paper include empirical evidence of the practices and challenges faced by organizations dealing with NFRs in CSE, and the implications we draw for research and industry. 
In addition, we bring awareness to academics and practitioners on the unique relationship, opportunities, and trade-offs that CSE offers to NFRs. 
In particular, 
\begin{itemize}
    \item \hll{how organizations using CSE can manage NFRs, for example by offloading sub-tasks of NFRs to third-parties,} %or leveraging metrics, the feedback loop, and continuous monitoring, where this opportunity of offloading also allows the organizations to ``realize'' an NFR without directly implementing it,
    \item managing an NFR comes at the cost of certain trade-offs, such as the loss of control over an offloaded NFR, or a decrease in the shared understanding of the NFR, and
    %\omar{Aren't trade-offs to be expected here? Why is this new?}
    \item the importance of \cm{} as an NFR and the amount of investment required to manage it.
\end{itemize}
\hll{We consider our empirically-derived insights to be useful hypotheses for future research to validate in more organizations that practice CSE.} The recent systematic literature review (SLR) by Yu et al. \cite{yu2020utilising} on the state-of-the-art (and practice) of leveraging continuous integration\footnote{Continuous integration is one aspect of continuous software engineering.} (CI) for NFR testing found that CI does indeed support the testing of \hll{some} NFRs, while highlighting the very low ratio of industrial empirical evidence to theoretical research in this area. 
Our findings bring empirical evidence on a broader set of practices in managing NFRs (including testing) in CSE practices.
%With the scarce empirical evidence on the challenges of handling NFRs in organizations practicing CSE, 
%Our contributions represent insights that are significant for both future research and software development practice. 
%We remind the reader that CSE practices are still emerging and our study is timely, as it guides our community's research directions by pointing to the need for empirical studies to investigate these implications as a result of these relevant opportunities from CSE.

% OLD CONTRIBUTIONS
% We also discuss a number of opportunities and trade-offs that these organizations encountered: 
% \begin{itemize}
% \item small organizations can offload the handling of some NFRs to third party providers but that comes at a cost of losing control over any offloaded NFR,
% \item CSE offer a tighter feedback loop and a more systematic approach to verifying NFRs, but the non-auto\-mat\-able NFRs lengthen the feedback loop, resulting in an organization not capitalizing on the benefits of CSE,
% \item configuration management practices in a CSE offer the ability to easily manage infrastructure and facilitate the implementation of other NFRs, but result in additional time investment in configuration and more software to maintain.
% \end{itemize}
%These insights help us understand the impact of handling NFRs in small organizations implementing CSE, and allow us to draw implications for further research and practice. 

In the remainder of the paper, we first introduce the related literature on NFRs and requirements engineering in agile environments, and on NFRs in CSE in particular. 
We then describe our empirical research methodology. 
We introduce our findings in the form of practices and challenges we identified at the three organizations we studied. 
Finally, our discussion of these findings debates the opportunities, but also the trade-offs associated with managing NFRs in organizations practicing CSE.   

\section{Background and Related Work}
Research into the ways in which CSE deals with NFRs is limited. 
We have a reasonably clear understanding of NFRs in general, and how agile software development and requirements engineering affect one another. 
However, these studies tend to be focused on larger contexts (e.g., in distributed teams or multi-team agile initiatives \cite{Alsaqaf2019}), and make little mention of CSE. 
In particular, the relatively recent rise of CSE \cite{Fitzgerald2014}, such as automated verification and build-on-commit, has the potential to greatly impact how NFRs are managed, because they insist on automated verification, rapid iteration, and shared codebases.
%\omar{Haven't shared codebases been around for a long time? I wouldn't say that's exclusive to CSE.}
We discuss related work in the areas of requirements engineering (RE) in agile settings, including NFRs, and finally, work on NFRs and CSE.

\subsection{Requirements Engineering and Agile}
It is now accepted that RE in agile organizations follows a just-in-time requirements engineering approach \cite{aranda2007requirements, Cao2008}.
Just-in-time RE practices deal with requirements as needed, rather than upfront, for example, by adding issues to the backlog and then delving into the requirements for that issue only once (or if) it becomes part of the iteration plan \cite{Ernst:2012wf}. 
Frequently members of an organization's leadership team are the only people with detailed knowledge of the requirements \cite{aranda2007requirements}, which aligns with philosophies such as Ries' lean startup approach~\cite{riesStartup}. 
The focus is to release often, gather feedback on the new features delivered, and prioritize work for the next release as needed. 
The implication of this just-in-time RE is that a) little upfront analysis is done and b) requirements analysis is taking place in the verification and experimentation phases, usually once software is released to customers. 
This has implications for how an organization is measuring and analyzing its requirements. 
Agile RE risks neglect of NFRs, since user stories focus primarily on features \cite{Inayat2015}.
Given that NFRs are difficult to analyze and understand, even in highly planned, up-front RE processes, suggests an even bigger challenge in just-in-time settings. 
%In the three organizations we study, agile practices included weekly sprint planning meetings, cross-functional roles, short iterations, and rapid releases.

For the purposes of our study we use Martin Glinz's definition of an NFR, which is ``an attribute of, or a constraint on, a system'' \cite{glinz}. 
\hll{On the surface, NFRs are often simplified as system qualities, such as the `ilities': usability, reliability, maintainability, etc.; however, upon deeper analysis NFRs may have significant influence on a system's overall design and architecture \cite{bellomo2014}.}
NFRs tend to receive a lot of attention in safety-critical systems, or in large organizations.
However, for small organizations in web application settings, including the organizations we studied, NFRs such as software reliability or system performance are also critically important (as we will show; see Table \ref{tab:nfr-ranking}). 
%\omar{There is no contradiction between the previous two sentences. Why "however"?}

Despite this importance, NFRs for organizations in agile settings are often ``informally stated, contradictory, difficult to enforce during development, and very hard to validate'' \cite{borg2003}. 
For example, 75\% of NFRs in a recent study from Eckhardt, Vogelsang, and Mendez \cite{Eckhardt:2016} were actually describing system behaviour . 
Finally, the wide-ranging and extensive NaPiRE study \cite{Fernandez2018} found that ``unclear / unmeasurable non-functional requirements'' were one of the top problems respondents had with requirements in their small organizations \cite[p.11]{Wagner:2017aa}. 

Moreover, \emph{how} and \emph{who} verifies NFRs are another important aspect of NFRs. 
Previous work found that NFRs are often difficult to verify and validate \cite{Nuseibeh2000,borg2003}.
Manual verification is often the most common choice to verify NFRs \cite{caracciolo2014software}. 
The study by Ramesh, Cao and, Baskerville \cite{Ramesh2010} suggested that in agile RE, NFRs often get de-prioritized in small organization settings and NFRs are ill-defined and ignored, e.g. ``We have no specific test of stability. We just test for functionality and see if it stays up'' \cite{Ramesh2010}. 
This makes dealing with NFRs at a later date more difficult, and it typically introduces technical debt.
% We heard similar issues in our interviews. 
% Our paper goes further than \cite{Ramesh2010}'s work by investigating how verification of NFRs is being addressed. 

Most recently, research has examined how best to incorporate NFRs into agile settings. 
For example Alsaqaf, Daneva, and Wieringa \cite{Alsaqaf2019} examined the way large, multi-team projects managed NFRs. 
The main challenge was the way NFRs cross-cut teams. 
Their research however studied organizations that are not operating in CSE, and which is the focus of the research we report in this paper. 

\subsection{Non-Functional Requirements and Continuous Software Engineering}
While there are other definitions of CSE, in this paper we use Martin Fowler's influential blog post on the topic \cite{fowler}. 
%\omar{Why do we use Fowler's practices? Why not someone else's?}
Fowler's definition of CSE includes maintaining a single source code repository, automated verification, automated builds, fast builds, and automated deployment \cite{fowler}. 
Fowler also specifies sound organizational practices, which include each developer creating a commit at least once a day and keeping the build process transparent for all stakeholders \cite{fowler}.
CSE is a paradigm that emphasizes rapid and automated releases of working software \cite{fitzgerald2017continuous}.

Research into the treatment of NFRs in CSE is limited. 
Similar to agile, it is mostly driven by FRs, with the primary focus of delivering functionality to end-users as early as possible to receive quick feedback \cite{bosch2014continuous}, often at the expense of other aspects such as NFRs.
While CSE has been shown to enhance requirement traceability \cite{Sthl2016}, the majority of research that investigates NFRs in CSE mostly focuses on verification of NFRs. 
%Chen \cite{chen_overcome_2017} wrote that the verification of NFRs in the context gaof CSE was an important aspect for future study. 
Although, NFRs are often neither comprehensively verified nor automated. 

One study on verification found that an organization may not provide sufficient time to verify NFRs \cite{Petersen2010TheEO} because NFRs may require more time to verify. 
Late verification of NFRs may cause severe side effects, such as re-factoring architecture at a stage when an organization is busy preparing for release of a software \cite{nilsson_visualizing_2014}. 
Even when organizations do verify NFRs, the amount of automated tests for NFRs is limited, and an organization may require manual verification to verify them \cite{leppanen_highways_2015}. 

%\omar{The next two paragraphs suddenly switches to CI instead of CSE. Why?}
One aspect often missing in those prior works is description of tools used for NFRs. 
For example, Savor et al. \cite{savor_continuous_2016} mention that NFRs were watched by measurement tools, but makes no mention of what tools or how the tools were operated. 
The recent SLR by Yu et al. \cite{yu2020utilising} found that CI environments (and tools) could be leveraged to adequately verify NFRs; however, they are underutilized.
Furthermore, there is a very low ratio of industrial studies in NFR verification in CI compared to academic studies \cite{yu2020utilising}.

While this previous research indicates that CI may be leveraged to \emph{verify} an NFR \cite{yu2020utilising}, verifying an NFR is only one important aspect of NFRs.
\hll{In particular, an organization may attempt to verify an important NFR to determine whether that NFR may or may not be realized, and perhaps to what degree.
However, verifying an NFR does not help actually realize the NFR itself.}

NFRs or quality attributes are typically decomposed into smaller units, either quality attribute scenarios \cite{barbacci2003quality} or tasks \cite{yu1997towards}. %(Yu, 1997). 
Realizing the NFR as a whole (or `satisficing' \cite{simon1956rational}) requires verifying that each of these smaller units is achieved.
Hence an NFR may be only partially realized, as any number of outstanding (or unknown) tasks may remain to fully realize (satisfice) an NFR.
%A previously realized NFR may degrade to the point of no longer being realized, unbeknownst to the organization due to the cross cutting nature of NFRs.
The type and number of these smaller units is context-dependent, as is the importance of each NFR. 
For example, one of our companies prioritized \nfr{scalability}, which they realized in part by offloading reliability tasks onto a cloud provider \cite{pahl2018architectural}.
% \colin{R1C1a needs work}
% \hghlt{Furthermore, implementing a new feature could negatively affect an NFR...?}
%Alternatively, an organization may chose to measure an NFR through metrics and set a minimum and/or desired levels or thresholds for each metric.
It remains unclear exactly \emph{how} CSE can help an organization realize an NFR.
Our study attempts to fill this gap by describing how organizations can manage and realize NFRs when using CSE. 

\begin{table}
\centering
\caption{Participants and their Roles at the three studied organizations}
\label{tab:participant-information}
    % \begin{tabular}{cccccc|cccccccc|ccccc} \toprule
    % ~         & \multicolumn{5}{c}{Alpha} & \multicolumn{8}{c}{Beta} & \multicolumn{5}{c}{Gamma} \\ \midrule
    % ~ & P1 & P2 & P3 & P4 & P5 &
    % P6 & P7 & P8 & P9 & P10 & P11 & P12 & P13 &
    % P14 & P15 & P16 & P17 & P18 \\
    % Developer & 
    % $\bullet$ & $\bullet$ & ~ & ~ & ~ & 
    % $\bullet$  & ~  & ~ & $\bullet$ & $\bullet$~ & $\bullet$~   & $\bullet$~ & $\bullet$~ & 
    % $\bullet$ & ~ & $\bullet$ & $\bullet$ & $\bullet$ \\
    % \midrule
    % Manager   & 
    % ~ & ~ & $\bullet$~ & $\bullet$~ & $\bullet$~ &
    % ~ & $\bullet$~ & $\bullet$~ & ~ & ~ & ~ & ~ & ~ & 
    % ~ & $\bullet$~ & ~ & ~ & ~ \\ 
    % \bottomrule

%\begin{tabular}{cclll}
\begin{tabular}{crlcm{.9cm}m{.8cm}}
\hline
\textbf{Org.} & \textbf{P\#} & \textbf{Role} & \textbf{Gender} & \textbf{Exp. at Org.} & \textbf{Overall Exp.} \\ \hline
\multirow{5}{*}{Alpha} 
& P1 & Dev. & Male & \textless~2y & \textless~20y \\
& P2 & Dev. & Male & \textless~10y & \textless~20y \\
& P3 & Mgr. & Male & \textless~10y & \textless~20y \\
& P4 & Mgr. & Male &\textless~5y & \textless~10y \\
& P5 & Mgr. & Male & \textless~10y & \textless~20y \\
\hline
\multirow{8}{*}{Beta} 
& P6 & Dev. & Male & \textless~2y & \textless~20y \\
& P7 & Mgr. & Female & \textless~5y & \textless~20y \\
& P8 & Mgr. & Female & \textless~10y & \textless~10y \\
& P9 & Dev. & Male & \textless~5y & \textless~5y \\
& P10 & Dev. & Male & \textless~5y & \textless~20y \\
& P11 & Dev. & Male & \textless~2y & \textless~20y \\
& P12 & Dev. & Female & \textless~2y & \textless~2y \\
& P13 & Dev. & Male & \textless~2y & \textless~5y \\
\hline
\multirow{5}{*}{Gamma} 
& P14 & Dev. & Male & \textless~2y & \textless~2y \\
& P15 & Mgr. & Female & \textless~2y & \textless~20y \\
& P16 & Dev. & Male & \textless~2y & \textless~5y \\
& P17 & Dev. & Male & \textless~5y & \textgreater~20y \\
& P18 & Dev. & Female & \textless~2y & \textless~5y \\
\hline
\end{tabular}
\end{table}

% \begin{table*}
% \centering
%     \begin{tabular}{cccccc|cccccccc|ccccc} \toprule
%     ~         & \multicolumn{5}{c}{Alpha} & \multicolumn{8}{c}{Beta} & \multicolumn{5}{c}{Gamma} \\ \midrule
%     ~ & P1 & P2 & P3 & P4 & P5 &
%     P6 & P7 & P8 & P9 & P10 & P11 & P12 & P13 &
%     P14 & P15 & P16 & P17 & P18 \\
%     Developer & 
%     $\bullet$ & $\bullet$ & ~ & ~ & ~ & 
%     $\bullet$  & ~  & ~ & $\bullet$ & $\bullet$~ & $\bullet$~   & $\bullet$~ & $\bullet$~ & 
%     $\bullet$ & ~ & $\bullet$ & $\bullet$ & $\bullet$ \\
%     \midrule
%     Manager   & 
%     ~ & ~ & $\bullet$~ & $\bullet$~ & $\bullet$~ &
%     ~ & $\bullet$~ & $\bullet$~ & ~ & ~ & ~ & ~ & ~ & 
%     ~ & $\bullet$~ & ~ & ~ & ~ \\ 
%     \bottomrule
%     \end{tabular}
% \end{table*}
\section{Research Method}
\label{method}
\begin{figure}[ht]
    \centering
    \includegraphics[width=0.45\textwidth]{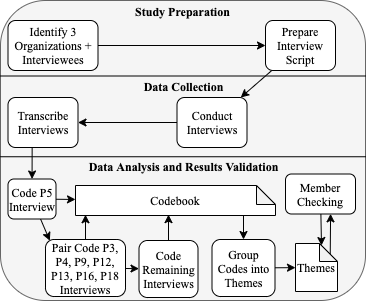}
    \caption{Research Process}
    \label{fig:research-process}
\end{figure}

\hll{We use an exploratory approach in a multiple-case study \cite{yin_case_2002} to investigate the practices in management of NFRs at three software development organizations using CSE.}
\hll{A case study methodology is the most appropriate research methodology when studying a contemporary phenomena in a real-life context, such as we are \cite{yin_case_2002}.
Given the intricate nature of this understudied research domain we employed a qualitative research methodology, which is suitable to study non-technical aspects, including socio-technical, that complement traditional quantitative software engineering research methods \cite{seaman1999qualitative} and to develop empirically-driven theories in software engineering \cite{easterbrook_selecting_2008}.}
%Thus, we utilized an exploratory approach in a multi-case study through qualitative research to investigate the practices (and challenges) faced by software organizations when managing NFRs in a CSE context.}
%Our methodology can be broken down into three phases: 1) study preparation, 2) data collection, and 3) data analysis and results validation.
Our methodology is summarized in Figure~\ref{fig:research-process}.

%\subsection{Study Preparation}
To recruit prospective organizations we used personal contacts at organizations using CSE, which resulted in identifying seven local organizations. 
We approached each organization to gauge their willingness to participate in our study and to understand the extent to which they manage NFRs and practice CSE in their development.
% \colin{why did we select three? to address R3C1}
% \zane{After initial engagement, several organizations dropped off due to a few factors, which primarily included degree of CSE practiced in organization and time constraints of the organization to commit to the study. We ended up studying three organization that covered....}
% \hghlt{something more??? make up some bullshit? cite SIGSOFT Empirical Standards \cite{ralph2020acm}}
\hghlt{Our prospective organizations covered a broad range of software domains (e.g. e-commerce, content management), number of employees, age of organization, and maturity of CSE practices. We selected three organizations based on availability and willingness to grant us research access, while each organization represented a different software domain and exhibited mature CSE practices for our study.}%and more importantly represented best-of-their-class with respect to managing NFRs.

In particular, each organization has grown from a small startup into a mature, established leader in their respective business domains, and is implementing the recommended CSE best-practices to a high degree.
% \colin{This should address R3C2?}
\hghlt{In addition, in a subsequent survey with the study participants, we found 92\% of respondents thought their respective organization manages NFRs well.}
In line with the tenets of case-study research, these organizations offered the opportunity to study the relationship between CSE and NFRs due to their realizing the importance of key NFRs and appropriately, and continuously, managing these NFRs early-on and throughout the organization's lifespan.
%If these organizations did not manage these key NFRs, then they would not be operating successfully today.
%\omar{Is the previous sentence a hypothesis, a statement based on literature, or conjecture? Because they're doing well even though some NFRs are not well managed.}
%Finally, each of these organizations is viewed as a leader of their respective niches; as such, each of these organizations enable us to learn from the best of the best.}
Each of these organizations are 8 years old, have between 30-60 employees, and use some form of cloud provider (e.g Amazon Web Services, Google Cloud).
Alpha, Beta, and Gamma all run automated builds, use automated tests, and use automatic deployments to production.

Alpha works in the %mobile [redacted]
data collection and analytics industry, processing large amounts of data on a daily basis.  %, and collects data from millions of users worldwide. 
Beta provides an e-commerce platform with multiple platforms for customers distributed worldwide.
Gamma is a content provider, including online advertisement management.
We chose these three organizations (Alpha, Beta, and Gamma) because our intent was to study how \emph{commercial organizations} perceived and managed NFRs in the context of CSE. 
\hll{As part of our approved %\footnote{University of Victoria's Human Research Ethics Board} 
research protocol and our NDAs, we anonymized the names of the organizations and interviewees.}
% put this in a footnote if it helps save us space.

\subsection{Preliminary Study}
\hll{We first sought to build our understanding by familiarizing ourselves with the personnel and the particular software development practices at each of the three organizations. 
One author spent multiple days over a few months embedded at each organization, learning about their products, processes, and lines of business \cite{potts_research}. 
While we did not have access to every single employee, through our immersive visits we were able to speak with a broad range of employees, spanning multiple organization boundaries.
In particular, we spoke with 37 different employees (representing roughly \( \frac{1}{3} \) of total staff), including 17 developers, 8 development managers, 5 product managers, 4 executives, 2 customer success specialists, 1 quality assurance member, and 1 director of sales.
%\neil{can we say what portion of the companies this nunmber covers?}
Thus we had access to a sub-set of employees, but included at least one employee from every major team.
Our early immersive meetings and observations at our organizations informed our focused investigation into the lack of shared understanding of NFRs and its relationship to rework.}

\subsection{Data Collection}
Once we had confidence in our understanding of each organization, we collected qualitative data through interviews of eighteen employees in development and managerial roles from these organizations
A summary of the interviewees is in Table~\ref{tab:participant-information}.
The semi-structured, open-ended interviews were conducted by two authors with each participant and lasted between 45-90 minutes in person at an organization's office.
We created a template of fourteen base interview questions for our semi-structured interviews\footnote{Full set of questions are in our research artifacts repository}.
Our interviews included questions on NFRs and CSE, such as: 1) organization definition of an NFR (e.g. \emph{Do you define NFRs? How do you define an NFR? Which NFRs are important to you organization?}) 2) organization definition and implementation of CSE (e.g. \emph{Are you familiar with the term continuous integration, delivery, and deployment? If familiar, how do you define those terms? Does your organization practice CSE?}), as well as about how their organization managed NFRs: 3) treatment of NFRs in context of CSE (e.g. \emph{How do you trace an NFR through the continuous deployment? What happens when an NFR fails; is there a feedback loop from continuous development?}). 

We started each interview by going through our base set of questions.
We followed with additional probing questions on more specific subjects, depending on an interviewee's role or primary work. 
For example, an interview with a DevOps engineer involved extended questioning on tools or processes used by the engineer's organization to facilitate their CSE. 
An interview with a front-end developer working extensively on the visual aesthetics of an application often included questions regarding prioritizing and verification of NFRs.
Interviews were, with permission, recorded.

\begin{table*}
\caption{Progression from raw text to code to theme. One snippet of raw text is related to one or more codes, and one theme relates to many codes.}
\label{tab:samplecode}
    \begin{tabular}{p{3.8in}p{1.6in}p{1in}}
    \toprule
    Raw Interview Text & Codes & Theme \\ \midrule
    ``We have a metric that tracks it to monitor it. So what happens is on a deployment I would monitor it and it is actually part of my responsibility to monitor over the day and see how that how those numbers are relative to the previous day.'' & \textsf{Metrics, Implicit, Deployment, Organizational, Manual,	NFRPerception, DevOpsPerception, Testing} & \emph{Put a Number on the NFR} \\
    \bottomrule
    \end{tabular}
\end{table*}

\subsection{Data Analysis and Results Validation}
We transcribed each recorded interview using an automated transcription service, and verified each transcription by a human.
Subsequently, we employed thematic analysis, \hll{an established data analysis method to identify themes and patterns in our data \cite{cruzes_recommended_2011}.}
%To acquaint ourselves with the interview data, we first read each interview, so that everyone gained familiarity with all interviews. 
%As each co-author did not attend every interview in person, reading each interview helped us gain familiarity and recognize initial patterns. 
Our analysis involved inductively developing codes from the raw transcripts and then identifying themes that related to practices and challenges of handling NFRs in our studied organizations. 
The open coding approach \cite{corbin1990grounded} was used to minimize the bias any one particular coder could have.
%we began to codify interviews as illustrated by Figure \ref{fig:research-process}. 
%To prevent any initial bias from affecting our choice of codes and potentially our themes, 
Throughout our coding we used the constant comparison method, whereby codes were added, removed, and merged based on the discussions between the coders. 

\hll{After each coding session all coders would meet to discuss each item, which codes were applied, and debate the reasoning behind a particular code and whether that code actually applied to that item.}
%These meetings allowed the coders to create a mutual understanding of exactly what each code meant and the outcome was that were only used codes that have been confirmed in a discussion.}
%We coded the interviews through an iterative, three-step process where at each step we used inter-rater agreement to develop and agree upon the codes in our interviews. 
In our initial coding phase, the first four authors coded every dialog segment of the same transcript (P5) (a `dialog segment' was our unit of analysis, and encompassed the interviewee's answer to a question) independently before calibrating with the other coders. 
As part of calibration, the first four authors conducted a mini-workshop to discuss their understanding of the codes after coding each transcript.
% \colin{This should address R1M4, no?}
\hghlt{We defined coder agreement as any dialog segment where all 4 coders had at least one code for the segment in common (we merged synonymous terms into  one  single  code).}
Since our mini workshops involved extensive discussion on the meaning and use of the codes in our codebook, the coders evolved a shared understanding of each code. 

%First, we used the longest interview transcript (P5, over 90 min of data) as a pilot to develop initial codes, a common understanding of codes, as well as a consistent way of coding the raw data. 
%This first transcript was split over three equal parts, where each part consisted of approximately 40 dialog segments (a `dialog segment' was our unit of analysis, and encompassed the interviewee's answer to a question). 

%As part of calibration, the first four authors conducted a mini-workshop to discuss their understanding of the codes after coding each part of P5's transcript.
%We defined coder agreement as any dialog segment where at least one code for the segment was common to all four coders (we merged synonymous terms into one single code). %The agreement levels improved from 16\% after coding the first part, to 50\% and 43\%  in the second and third part respectively; both results can be considered fair and demonstrated drastic improvements from the first part. 
%Since our mini workshops involved extensive discussion on the meaning and use of the codes in our codebook, the coders evolved to share understanding of each code. 
%Given that an agreement of 40\%-75\% is considered fair to good \cite{lwakatare_devops_2019}, we felt the improvement was significant and we continued with coding other interviews. 
 
Following the first phase of coding, we coded an additional seven interview transcripts distributed across the three organizations (P3, P4, P9, P12, P13, P16, and P18). 
The intention was to further develop codes and consistency in coding. 
\hll{In this stage, each interview was individually coded by two separate authors; inter-rater agreement levels varied from 64\% to 93\%, with an average agreement of 85\%.}

In the last phase of coding we divided the remaining interviews and assigned a coder for each interview.
% \colin{addresses R3m4}
\hghlt{The number of codes created in the last 10 interviews accounted for only 4 additional codes, thus our codebook was saturated.}
To ensure that a coder did not miss any important codes during individual coding, we included a sanity check step where another coder would check the first coder's results.
Table~\ref{tab:codebook} shows a sample of six codes from our codebook, while the entire codebook (61 codes) can be found at in our research artifacts. 

\begin{table}
\caption{Codebook Examples}
\label{tab:codebook}
\begin{tabular}{p{2.5cm}p{5.5cm}}
\toprule
\textbf{Code Name} & \textbf{Description} \\
\midrule
Configuration\-ManagementNFR & NFR: \cm{} \\
NFROffloading & Relinquishing technical control / responsibility of said NFR to an external entity \\
NFRPerception & Individual's perception of NFRs \\
PerformanceNFR & Performance of the outcome (ie the output) usually measured by time \\
Metrics & Creating or monitoring of a quantitative value \\
Tooling & ``Off the shelf'' (Kubernetes, Docker, etc) \\
\bottomrule
\end{tabular}
\end{table}

To answer RQ1 and RQ2 we developed themes based on thematic synthesis \cite{cruzes_recommended_2011} of our coded data.
% \colin{R3m3}
\hghlt{We discussed similarities and differences between codes to group codes into clusters \cite{boyatzis1998transforming}, whereby each cluster had a distinct higher-order theme.}
%All of our codes were grouped into themes by performing similarity analysis to identify codes that were often assigned together.
Each theme represented either a practice (RQ1) or a challenge (RQ2).
Table~\ref{tab:samplecode} shows an example of relating a raw transcript quote to an eventual theme.
Finally, to increase the credibility and validity of our research findings, we performed member checking \cite{MilesHuberman1984} with the study participants to verify whether our findings resonate with the context of their organization. The member checking feedback was used to revise our findings. 

%===============================================================================
\section{Study Results} \label{findings}
%shortened results considerably -- mostly moved to discussion
In this section we describe the themes (4 practices and 3 challenges) we derived from our analysis.
We summarize these practices and challenges in Table \ref{tab:findings}. 

%===============================================================================
\begin{table}[bh]
\caption{Practices and Challenges of Organizations Handling NFRs with CSE}
\label{tab:findings}
    \begin{tabular}{cl}
    \toprule
    \multirow{4}{*}{Practices}& Put a number on the NFR \\
    & Let someone else manage the NFR \\
    & Write your own tool to check the NFR \\
    & Put the NFR in source control \\
    \midrule
    \multirow{3}{*}{Challenges}& Not all NFRs are easy to automate \\
    & Functional requirements get prioritized over NFRs \\
    & Lack of shared understanding of an NFR \\
    \bottomrule
    \end{tabular}
\end{table}
%===============================================================================

All three software organizations in our study care deeply about NFRs in their software development. 
Alpha, in the data business, is most concerned about \cm{}, \nfr{security}, and \nfr{scalability}. These three NFRs are vital to Alpha's business as Alpha processes and stores data from large numbers of users per day on a third-party hosted infrastructure.
Beta, a leading e-commerce platform, is primarily concerned with \nfr{usability}, \nfr{performance}, and \nfr{stability/reliability}, as their applications handle millions of commercial transactions per day.
Finally, Gamma, an online advertisement content provider, requires \nfr{performance}, \nfr{revenue}, and \cm{} to ensure that Gamma's infrastructure can effectively deliver content to a wide audience.

Table \ref{tab:nfr-ranking} shows the overall ranking and individual ranking for each organization based on the frequency that the code representing the NFR appeared in our data (with at least 30 occurrences). The entire codebook can be found in our research artifacts.
Gamma's number 2 ranking (\nfr{revenue}) is notably absent from the table due to less than 30 overall occurrences, despite the high number of occurrences at Gamma.

Throughout our paper (and in Table \ref{tab:nfr-ranking}), we link our codes or particular findings to the evidence (interview text) as much as possible using quotes as well as spark-histograms. 
First introduced by Ying and Robillard \cite{Ying2014}, these histograms represent a compact and quick form of assessment of our findings: each histogram captures the 18 participants interviewed on the x-axis, while the y-axis shows the number of times the given code was mentioned by each participant, relative to the total number of mentions of that code (i.e. normalized). 
We also show the total number of mentions following the histogram.

The x-axis is ordered to match Table \ref{tab:participant-information}.
For example, consider the histogram for our code \cm{} \histno{ConfigurationManagementNFR}{103}, which shows participants 3-6 mentioned this code more often than the other participants, and across all transcripts, this code occurred 103 times.  
%\omar{The histograms do not show up well when printed. Also, this might be just me, but I can't make out which bar maps to which participant.}

%===============================================================================
\begin{table}
\caption{Ranking of NFRs by frequency of mention. Sorted by overall frequency. Columns Alpha, Beta, \& Gamma refer to rank within each organization. (T indicates a tie)}
\label{tab:nfr-ranking}
\begin{threeparttable}
\begin{tabular}{rlcrrr}
\toprule
\textbf{Rank} & \textbf{NFR} & \textbf{Histo.} & \textbf{Alpha} & \textbf{Beta} & \textbf{Gamma} \\ \midrule 
1 & Configurability & \histogram{histogram/ConfigurationManagementNFR.tex} & 1 & \tied{12} & 3 \\
2 & Performance & \histogram{histogram/PerformanceNFR.tex} & \tied{4} & 2 & 1 \\
3 & Security & \histogram{histogram/SecurityNFR.tex} & 2 & \tied{5} & 6 \\
4 & Scalability & \histogram{histogram/ScalabilityNFR.tex} & 3 & \tied{5} & 5 \\
5 & Usability & \histogram{histogram/UsabilityNFR.tex} & \tied{14} & 1 & \tied{9} \\
6 & Reproducibility & \histogram{histogram/ReproducibilityNFR.tex} & \tied{4} & \tied{5} & 4 \\
7 & Testability & \histogram{histogram/TestabilityNFR.tex} & \tied{6} & 4 & 8 \\
8 & Stab./Reli. & \histogram{histogram/StabilityReliabilityNFR.tex} & 10 & 3 & \tied{9} \\
9 & Availability & \histogram{histogram/AvailabilityNFR.tex} & \tied{6} & \tied{5} & \tied{12} \\
10 & Maintainability & \histogram{histogram/MaintainabilityNFR.tex} & 8 & \tied{5} & \tied{17} \\
11 & Readability & \histogram{histogram/ReadabilityNFR.tex} & 9 & \tied{15} & \tied{14} \\
\bottomrule
\end{tabular}
%\begin{tablenotes}\footnotesize
%\item[*] T indicates a tie.
%\item[] \-bility
%\end{tablenotes}
\end{threeparttable}
\end{table}
%===============================================================================

%===============================================================================
\subsection{Practices For Handling Non-Functional Requirements in Continuous Software Engineering (RQ1)}
%===============================================================================
Our first research question examines how NFRs are managed within each organization---the practices they use to handle NFRs in the context of agile teams following CSE. 
We identified four practices from our analysis. 
These were: \emph{Put a number on the NFR, Let someone else manage the NFR, Write your own tool to check the NFR}, and \emph{Put the NFR in source control}. 
We discuss each practice in turn. 

%===============================================================================
\subsubsection{Put a Number on the NFR}\label{sec:res_metrics}
%===============================================================================
% Relate back to our challenges, (whether the practice overcomes the challenge or whether the challenge limits the practice
%===============================================================================
The first practice for dealing with NFRs is establishing metrics to help validate and assess a particular NFR.
The concept of NFR metrics was frequently discussed during our interviews \histno{Metrics}{154}.
When an organization tracks pertinent metrics for NFRs, metric indicators collected by the organization's deployment pipeline provide valuable insight on NFRs.

For Gamma, \nfr{performance} is the most important NFR and is primarily tracked through a \q{[caching ratio] that we have from [redacted] our caching provider goes [on a dashboard] because that's usually a pretty good indication that something has gone wrong. It also drastically affects \nfr{performance}} (P14).
%However, Gamma could also evaluate \nfr{performance} with the addition of other metrics: \q{Some of the metrics [we use], for example revenue or [number of users] a specific page has had [...], might be able to infer \nfr{performance} of the website basically based on how many users they thought versus [how many actually used the page]} (P18). 
%For Gamma, \nfr{performance} and \nfr{availability} are important aspects of the organization's business.
%Since Gamma's \nfr{revenue} is driven by advertisements, a failure to load their application or a ``laggy" application experience would drive users away. Hence, \nfr{performance} ranks first and \nfr{availability} ranks seventh by frequency for Gamma as shown by Table \ref{tab:nfr-ranking}.
%Alpha collects data from millions of users worldwide
At Alpha, a drop in response times below a specific metric will cause a decline in \nfr{performance}. 
As a result, they \q{have a certain amount of monitoring set up [...] You're also defining which alarms are set. Therefore, if requests [drop] below [redacted] milliseconds and that's what you want to hold it to. Then that would be codified in the alarm} (P5).
Without setting a quantitative metric, Alpha may not receive warning that its software experienced a dramatic drop in \nfr{performance}. 

Similarly, Beta provides a platform for many retailers and \nfr{usability} is important for those retailers.
\hll{As part of managing \nfr{usability}, Beta tracks the number of customer actions required to complete a transaction through the use of \nfr{usability} metrics: \q{I can tell you that 90\% of our customers have less than [redacted] items [...] They'll [say] we know that each [order] requires [redacted] page loads} (P7).
While this may only represent a subset of tasks for \nfr{usability}, Beta views it as part of `satisficing' \nfr{usability}, in particular by bringing awareness to other teams.}
\hll{While at Gamma metrics are a key component in managing \nfr{usability}, across cross-functional teams, including development and product management by \q{showing how many users are using a specific feature, where the feedback benefits developers but our product team in their ability to make their decisions} (P16).}
%\hll{While neither of these examples can conclusively represent the entity of \nfr{usability}, the respective organizations determine this represents satisfying these NFRs.}

A critical success factor putting a number on the NFR is the feedback loop \histno{FeedbackLoop}{46}{} to enable the continuous monitoring of metrics.
The feedback loop is an integral part of CSE \cite{fitzgerald2017continuous} and is one of the earliest perceived benefits of adopting CSE.
For Alpha, which deals with a lot of user data, effectively managing \nfr{security} awareness is important for its business, e.g. \q{at least people who need to be aware of IAM changes are automatically notified} (P3). 
%Hence, the presence of the feedback loop enhanced with automatic notifications regarding infrastructure \nfr{security} positively impacts developer workflow.
%\q{A few things that we've got various forms of monitoring [...] For instance, when people are changing IAM policies not that there's automated feedback on that, but at least people who need to be aware of IAM changes are automatically notified} (P3).
%As Beta operates an e-commerce platform that serves many stores, it is imperative that the platform is available to customers. 
At Beta, the feedback loop provides the ability to quickly identify bugs that crash the software: \q{I think quick feedback is the core of DevOps so that we will see what broke} 
% or will see when [software] breaks quickly} 
 (P6) and 
% \q{by being alerted of things that aren't working sooner rather than later. It allows us to move much quicker much more quickly and our development and 
 \q{it reduces risk because things are integrated more frequently} (P6).
In particular, the feedback loop is most useful if it is a \emph{quick} feedback loop and many organizations strive to reduce the time required for a feedback loop to complete, e.g. \q{developers really want [the] feedback loop to be tight} (P13). 
 
%===============================================================================
\subsubsection{Let Someone Else Manage the NFR}
%===============================================================================
% Relate back to our challenges, (whether the practice overcomes the challenge or whether the challenge limits the practice
%===============================================================================
The most popular approach to managing NFRs we found was to let someone else do it by offloading it \histno{NFROffloading}{155}{}, where the `someone else' is typically a large cloud-service provider. 
Offloading an NFR means that an organization allows another platform or tool \histno{Tooling}{262}{} to realize, \hll{at least part of,} the NFR on the organization's behalf.

\hll{An organization's ability to focus on core functionalities and behaviour of their software is heightened by offloading the brunt of the work to realize NFRs such as \nfr{scalability}, \nfr{reliability}, etc. 
However, offloading an NFR is not as simple as flipping a switch to realize an NFR.
In many cases some form of configuration is required; furthermore, ensuring the system is ready for configuration, i.e., is configurable, requires a specialized skill set.}

\cm{} was the most referenced NFR \histno{ConfigurationManagementNFR}{103}{}. 
We found that the organizations we studied made frequent use of \cm{} tooling such as Docker, Terraform, and Kubernetes.
These tools help Beta maintain three separate software stacks through dependency management, which allows developers and testers to easily create or re-create an environment and application, e.g. \q{it automatically does it in the Docker compose files. So we've automated a lot of dependency updates and stuff like that} (P11).
Furthermore, cloud providers are favoured by DevOps engineers as they have access to a cloud provider support team to assist with issues, \q{cloud providers are awesome. I love being able to just file a support ticket} (P12), as opposed to dealing with an issue on their own.
Finally, \cm{} tools also help manage \nfr{reproducibility} \histno{ReproducibilityNFR}{52}{}, e.g. \q{the most important part is that the builds and the environment they run on and deploy to are defined in a repeatable way or state} (P5).
%and \q{I think some people believe Docker helps extract abstract complexities and difficulties and provides benefits that outweigh, its learning curve and overhead costs} (P1).
%\cm{} is an emerging NFR that refers to the need to ensure the code and associated hardware are in a sound state at any point in time. 
%However, many organizations do not manage their own hardware or infrastructure.
%Instead, they rely on cloud providers, such as Amazon Web Services (AWS), Microsoft Azure, or Google Cloud, for their infrastructure needs.

% security 
\nfr{security} \histno{SecurityNFR}{68}{} was also offloaded to a third-party service. 
\nfr{Security} and \nfr{privacy} checks may be intrinsically supported by a third-party service, reliving an organization of the obligation of laboriously maintaining and provisioning these checks on its resources. 
%Beta maintains an extensive customer database and are keenly interested in \nfr{security} and \nfr{privacy} of the database.
\nfr{Security} aspects can be offloaded, e.g. \q{I believe [\nfr{security}]'s all codified in like the Docker and Kubernetes world} (P6). 
%For instance, Kubernetes can use namespaces to isolate user access control. Other tools can scan for malicious third-party package dependencies.
Gamma utilizes a security key management system directly from one of the cloud providers.
%, e.g. \q{we use a security Magic Key Management pool from [redacted]} (P16), which frees Gamma developers from having to worry (extensively) about key management issues.

% \nfr{performance} & \nfr{availability} & \nfr{scalability} & stability/reliability
Cloud providers are heavily relied upon to manage \nfr{performance} \histno{PerformanceNFR}{88}{}, \nfr{availability} \histno{AvailabilityNFR}{39}{}, \nfr{stability/reliability} \histno{StabilityReliabilityNFR}{41}{}, and \nfr{scalability} \histno{ScalabilityNFR}{56}{}.
For example, at Alpha, \q{[your web application is] immediately spread across however many \nfr{availability} zones and nodes as you want} (P5).
% and \q{we are concerned with \nfr{performance} [...] latency, \nfr{scalability}.} (P3)

Finally, NFR offloading provides the capability to ratchet NFRs \cite{bellomo2014}, as all three organizations highlighted their ability to increase the amount of resources they consume from their cloud providers if they hit a particular NFR bottleneck. 
%Increasing the amount of resources obviously has financial cost.
%These additional resources allow the organization time to properly investigate the source of the bottleneck.
For example, Gamma and Alpha both noted that they can simply pay more to the cloud provider, e.g., 
%\q{something's going burning through too much. We gotta deal with it now versus hey is burning too much and it'll cost a few bucks we can deal with it tomorrow} (P17) and 
\q{We just put money in the machine and made it better} (P5).

%===============================================================================
\subsubsection{Write Your Own Tool to Check the NFR}
\label{sec:find_own_tool}
%===============================================================================
% Relate back to our challenges, (whether the practice overcomes the challenge or whether the challenge limits the practice
%===============================================================================
As opposed to offloading an NFR to a third-party service or tool, organizations also wrote custom, in-house source code, custom tooling, scripts, or manifests \histno{ImplOfNFR}{78}{}.
%For example, \cm{} was facilitated by utilizing Terraform\footnote{Terraform is an automated cloud deployment scripting language} scripts, which then largely handled deployment and operations concerns.
%, you document some infrastructure design decisions as code, which are then an integral part of the continuous pipeline and the importance of getting them right the first time, \q{a lot of those design considerations and non-functional requirements that go on, am I going to talk about how was the response time for example that dictates initial architecture and that has a huge impact. So you end up, if you're lucky you end up with a trivial use case. It's like, oh that's what that's designed for. And that's what that's designed for. You only do it once. You have to really worry about the \nfr{scalability} because it works out}. This is important to company Alpha because their business model relies on them being able to handle a large amount of data
Beta and Gamma both codified some \nfr{usability} parameters of their user facing front-end \hll{to manage a portion of \nfr{usability} according to their individual definition and satisfaction, which may not be broadly applicable.}
This codification was done as part of their source code. 
They submit the source code, wait for the source to process through the CSE pipeline, and finally observe and verify the result, \q{this is how we define \nfr{usability} and make sure that it's there. 
If you want to change our \nfr{usability} parameters or whatever we change it in source code and then we can test it and verify that it still meets our needs} (P10).

%Custom tooling (~\histogram{histogram/customTooling.tex}~) was discussed by only eight interviewees, thus much less prevalent than third-party tools.
For a resource-constrained organization a custom tool is usually a last resort, where the existing off-the-shelf solutions either do not exist or do not sufficiently meet particular requirements.
A custom tool may be based on an augmented third-party tool that requires significant modification to meet the specified needs, e.g. \q{so we used to have [name redacted] dashboards out there ... [but that] didn't give us any application specific information} (P14).

%. But I think that we were feeling that we weren't getting enough customization out of it because it was just the metric [redacted] was giving us directly of the instances and 

Some custom tools were used to handle NFRs from an operations perspective to determine \nfr{availability} \histno{AvailabilityNFR}{39}{} or \nfr{stability/reliability} \histno{StabilityReliabilityNFR}{41}{} of the infrastructure. %, e.g. \q{I think we have a script running that gets health checks of our servers, and I think it's automatically knowing if something goes down---if they're not getting that health check then they're posting to our channel and then the appropriate person will get notified and they'll put those servers back up} (P11).
Other custom tools were developed to help automatically enforce or validate \nfr{security} \histno{SecurityNFR}{68}{} within a CSE pipeline: \q{so there's a lambda [function] that runs when you make a bucket, it triggers and goes 'you didn't encrypt', it turns on encryption, tells you, 'you are an idiot'. \nfr{Security} non-functional requirement!} (P4).

%===============================================================================
\subsubsection{Put the NFR in Source Control}
%===============================================================================
\label{sec:res_transfer}
%===============================================================================
% Relate back to our challenges, (whether the practice overcomes the challenge or whether the challenge limits the practice
%===============================================================================
%In a CSE context, communicating NFRs to one's team is vital.
At the organizations we studied, the workforce was constantly changing (typically growing), and the products were experiencing a rapid pace of change.
The constant change requires that developers gain a clear understanding of the NFRs of the product so they know how their changes might impact these NFRs. 
Typically, an NFR is not effectively communicated via natural language documents: \q{with respect to codifying something vs documenting it: it's not precise enough it's written in English. It's open to interpretation.} (P17)

Our results show developers captured NFRs directly in source control. 
`Codification' refers to using code and related artifacts (such as version-controlled JSON configuration files) to capture NFR knowledge \histno{Codification}{92}{}. 
For example, automated dashboards monitor health indicators, such as \nfr{availability}, and these explicit rules or triggers that represent metrics of an NFR are in source-control.
%Terraform\footnote{\url{terraform.com}} scripts capture NFRs related to deployment infrastructure, such as memory allocation.
Codification helps set an objective metric for developers who might not have had enough time to acquire the tacit knowledge about what constitutes an acceptable NFR threshold.
P17 notes \q{my understanding is that we should be able to see our tests in source control in terms of nicely capturing results in a way that I and the other developers can see and say some data is not captured yet} (P17).
%Other NFRs are directly captured in source code: \q{so it's in code in source. This is how we want all our pages to look this is how we want everything to function [...] That's documenting it essentially} (P10).

%===============================================================================
\subsection{Challenges (RQ2)}
%===============================================================================
While the organizations in our study have concrete practices for managing NFRs, they still face challenges \histno{Challenges}{71}{}.
%. 15 out of 18 of our interviewees experienced challenges when operationalizing NFRs 
The most often described challenges were \emph{difficulty using tools and tests with NFRs, difficulty prioritizing NFRs}, and \emph{challenges with knowledge transfer}.

\subsubsection{Not All NFRs are Easy to Automate}
Some NFRs, such as \nfr{usability}, are intrinsically difficult to verify through automated means. 
%, %e.g. \q{I have done some [testing] using [UI test tool] selenium. Selenium is not fun. There's only one way you can test those things, for a human [to] sit down} (P4). 
%: \q{I find testing UIs is a very difficult process} (P4). 
% In the sample quote, the interviewee voiced his concern towards verifying \nfr{usability} NFRs and subsequently voiced their frustration towards automated UI verification tools: 
Unfortunately, adopting CSE approaches to development, which means committing and deploying at an extremely rapid rate, appear to make this an even greater challenge. For example, Beta relies on some manual \nfr{usability} acceptance verification for any customer-facing software deployment. As a result the people tasked with manual verificationour findings suggest
of \nfr{usability} become a bottleneck in CSE:
\q{[User acceptance testing is] easier to chunk into one deployment rather than 12 a day.} (P7).
%[The deployments becomes] a bottleneck I think [at] times for us. From my perspective once maybe twice a day is probably a reasonable number [...Anything more] I feel like would be a little bit of a challenge on the operational side to keep up with that}} (P7).
% \nfr{usability} is hard to verify, as it is in other settings, but \nfr{usability} is especially difficult in SMEs where often only one person is in charge to verify \nfr{usability} if automated tests are unreasonable. 

In addition to \nfr{usability}, it may be difficult to write automated tests for other NFRs: 
%\q{We've tried a couple times but we haven't been very successful with the [tests]} (P16) and 
\q{There's a lot of variability and it's hard to write really good thresholds of what is working. What is not working} (P14).
Although some NFRs may not require much work to automate, based on the interviewees' sentiment, producing the ``right'' automated tests may require more than one test creation iteration. 

Furthermore, purely increasing the number of tests does not equate to higher quality tests: \q{I think that testing itself [is a] quality metric. So we're looking at coverage as a possible metric but we're trying to determine a more accurate form because we don't believe that [increasing] code coverage [will] provide us the value [...] if you test all cases that you have 100\% coverage, it's not really scalable} (P16). 

Under normal circumstances, it can be difficult to predict a sudden spike in user activity.
If Gamma suddenly experiences overwhelming levels of traffic, prior tests for \nfr{performance} and \nfr{scalability} may not suffice: \q{I feel like [tests] always [had] a bias toward the happy cases [...] When something does go wrong, it's something horribly out of left field [...] How do you prepare for those? How do you think about what left field is?} (P18).
Determining the parameters and conditions that would effectively verify the entire problem space of an NFR is difficult.

% A key concept of CSE is the rapid collection of feedback from customers to improve a software. 
% However, when an organization fails to effectively verify its NFRs or developers received minimal feedback from tests, the feedback loop facilitated by CSE cannot help treat NFRs: 
% %\q{But as a developer, I don't see much of that [\nfr{usability} and learnability] feedback unless it's gone through product. So 
% \q{[...] the only understanding of the \nfr{usability} of the page for me is how I [and other developers] use it} (P11).
% Without receiving verification and usage feedback, the interviewee cannot exert efforts to improve \nfr{usability} as desired by customers.
% Verifying NFRs represents an important aspect of an organization's adherence to a CSE and good RE practices. 
% Yet, based on our study, automating the verification of NFRs is not without challenges. 

%===============================================================================
\subsubsection{Functional Requirements Get Prioritized Over NFRs}
%===============================================================================
% Make the main point of this section that NFRs are easy to drop when resources are tight. 
% NFRs become a FR when NFRs become a problem.
Since our collaborating organizations are still growing rapidly, employees balance many other responsibilities, among other potential limits to resources. 
In resource-constrained environments, NFRs are an easy target to bump from the sprint plan or milestone due to lack of clarity around how to verify or define an NFR.
At Alpha, one aspect of their system is the need for CPU power to process large volumes of data. 
However, if not enough developers are available to maintain the system, \nfr{efficiency} and \nfr{performance} may degrade over time:
\q{nobody's looked at this last six months maybe someone should check it out [...] It's a resource management issue and a lot of times you have too many things for too few people.} % [People] kind of make things and then go on to the next thing} 
(P4). 
An organization may be obliged to make NFR trade-offs with the hope that immediate, short-term success will lead to the ability to remedy the trade-offs, i.e., potential technical debt, in the future. 
In our study, 14 out of 18 interviewees \histno{TradeOff}{41}{} acknowledged the existence or previous existence of trade-offs: \q{the sort of startup code today some of it you could consider clever but you do too many clever things then rack up a lot of technical debt} (P5). %and \q{It's mostly just down to just I call the law of raspberry jam. I phrase it as you can only spread things so thin before [it] doesn't cover everything} (P3). 
%In this instance, the interviewees implied that a startup organization may willingly make trade-offs for present success. For example, choosing \nfr{deployability} over \nfr{maintainability}, \nfr{performance} over stability, or \nfr{usability} over \nfr{readability}.

An organization, even an early-stage organization, needs to be aware of these NFR trade-offs so that it can ensure that an important NFR such as \nfr{scalability} is improved when the NFR reaches a low point: \q{Some of the core pieces of the system again get more love or more time to knock that [technical debt] number down or we just pay more close attention} (P5).
% [...] That's what always a constant battle between resources.} (P5). 
% an organization must be strategic when choosing prioritizing the issues to fix: 
% knowingly write code that may cause technical debt and problems in the long term, but in the interest of quickly delivering value to customers, an organization may temporarily accept the trade-off.
%Moreover, when an organization does intend to address its technical debt, limited resource may restrict the organization's resolve. 

% Fortunately, our interviewees for the most part recognized that limited resources is a challenge to operationalizing NFRs. Once an organization is able to direct more resources, this recognition may be act as a catalyst to alleviate these challenges. 
%===============================================================================

%===============================================================================
\subsubsection{Lack of Shared Understanding of an NFR}
A shared understanding of an NFR implies that everyone involved with the NFR is in agreement with  the meaning of that NFR, and its various components. 
This shared understanding relies on knowledge transfer from the people---such as the product manager or CEO---who created the NFR, to those whose work might affect it. 
Shared understanding is a challenge for our subject organizations.
They faced problems with inconsistency in what gets explicitly stored in source control; with tacit knowledge and a low (bad) circus factor \cite{Cosentino2015}; and problems with role siloing. 

Our study found organizations relied inconsistently on documentation for knowledge transfer of NFRs \histno{KnowledgeTransfer}{78}{}.
Explicit NFR knowledge transfer \histno{knowledgetransferANDexplicit}{55}{} occurs when a developer relies on a formal metric or artifact, such as documentation, to frame their understanding of whether an NFR is being met.
For instance, in reference to having their infrastructure run by Terraform scripts to deploy using Kubernetes, one of our respondents mentioned: 
%\q{aside from it being the new hotness? (laughs) It allows scaling in different ways. At least that's 
\q{a big incentive for me is the idea that I don't become a linchpin and at the same time a bottleneck being the one person specialized in this} (P3).
% or having a small team of people who are specialized in it, but that it centralized that knowledge which to me is important for scaling up things quickly}

Our subject organizations do not consistently invest effort or resources in documenting NFRs.
While some NFRs are being actively monitored (see section \ref{sec:res_metrics}) or are documented as code within a source control system (see section \ref{sec:res_transfer}), others are not.
For example, in reference to the fact that the \nfr{performance} of a feature requires processing to finish within 2.5 hours: \q{No, I don't think I have that specifically labeled. I don't think I outlined any specific requirement like that} (P2).

Implicit knowledge transfer of NFRs \histno{knowledgetransferANDimplicit}{29}{} occurs when someone on the development team attains a personal understanding of an NFR without relying on an established metric or artifact.
For example: \q{at the moment it is tacit knowledge and unfortunately when a new developer comes in and starts working on stuff [they struggle]} (P10). 
The `circus factor'\footnote{We avoid the ugly implications of `bus factor` in favor of circus factor: the number of people who have to run away to join the circus to hurt the project.}  \cite{Cosentino2015} captures a major problem with implicit knowledge: \q{I try not to get hit by a bus. So does [redacted]. Certain parts of the system are maintainable because certain people know how they work versus it being [explicitly] documented} (P5).

% Bus factor
Implicit knowledge is often obtained by one or two people who have the overarching view (typically early employees or founders).
\q{I know a lot of the team leads have a ton of non-functional requirements in their head and how things should work and they're kind of the ones gating what goes out based on those undocumented non-functional requirements. If we were to document those we could get those ideas into the heads of the people actually writing the code, and better the development experience} (P13). 
Without transferring the knowledge of NFRs to front line developers, awareness of the importance of particular NFRs is lost. 

% boundary object/silo of roles
While explicit knowledge may be in source control (see Section \ref{sec:res_transfer}), role siloing makes understanding the artifacts difficult \histno{SiloOfRoles}{51}.
For example, capturing NFRs in the deployment scripting language Terraform is likely highly useful for team members working closely with deployment and DevOps roles, e.g. \q{For anything that I do, [infrastructure as code] ends up in Terraform as a form of documentation} (P3). 
However, the value of documentation provided solely by code can vary depending on developer context. 
Developers less familiar with Terraform may have different interpretations of what the script is doing, if they can understand the Terraform language in the first place.
\section{Discussion} \label{discussion}
%===============================================================================
Our empirical study sought to unveil the state of practice in managing NFRs in organizations that use CSE. 
%NFRs, equally important as functional requirements and difficult to manage in software projects, bring additional hurdles in CSE where the focus is on automated and rapid release of software to deployment environments. 
In particular, we studied three organizations, each developing software with several vital NFRs and exhibiting mature continuous software engineering.
% \colin{address R3C3}
\hghlt{Our findings suggest that the studied organizations manage NFRs in CSE through four main practices.}
% \colin{address R3C2}
\hghlt{We believe these practices are best-practices for managing NFRs, as our respondents, overall\footnote{11/12 respondents believe NFRs are well managed}, are very satisfied with how their respective organization manages NFRs.}
While these practices may not be specific to CSE, we believe a special relationship exists for each that is unique in a CSE context.

While the use of metrics is well-established in industry, CSE enables an organization to better automate and deploy metrics in a rapid feedback loop.
Furthermore, while NFRs are typically more difficult to automate, CSE brings a heightened focus,  attention, and importance to automating important NFRs.
While the lack of shared understanding of NFRs has been commonplace, we found evidence to suggest the CSE has led to a decrease in shared understanding \cite{werner2020re}.
Finally, while FRs may be often prioritized over NFRs, the fast paced environment of CSE exacerbates the deprioritization of NFRs.
%Thus we believe that our findings on these practices and challenges, while not necessarily new, are important to consider in a CSE context.}

%Yu et al. conclude that CSE practices are underused for NFR verification \cite{yu2020utilising}.
%\todo[inline,color=red]{NE: do we agree underused?} YES, makes our results more meaningful.

Our results allow us to reflect on the somewhat surprising opportunities that CSE practices offer to managing NFRs, as well as associated challenges and trade-offs when managing NFRs in CSE.
We also discuss the importance of \cm{} as an NFR in organizations practicing CSE.
Our findings represent valuable empirical insights that add to the nascent empirical evidence on utilizing CSE to manage NFRs \cite{yu2020utilising}.
Finally, we propose research directions for the treatment of NFRs in CSE. 

%===============================================================================
\subsection{Non-Functional Requirements in Continuous Software Engineering: A Silver Lining}
%===============================================================================
NFRs often do not get the appropriate attention they deserve.
NFRs are cross-cutting in nature as they impact many aspects of the system and may be difficult to decompose into fragments that can be realized in a short, rapid CSE iteration \cite{bellomo_evolutionary_2014}, which further complicates the ability of an organization to manage an NFR.
%\hll{Even defining what it means to realize an NFR is non-binary and relative to each organizations' individual goal.}
However, our evidence suggests that it might be easier, for organizations that shift to CSE, to manage through one of the four practices we identified (``Put a Number On the NFR'', ``Let Someone Else Manage the NFR'', ``Write Your Own Tool to Check the NFR'', and ``Put the NFR in Source Control'').
%\omar{Didn't we make the case that there were trade-offs in the previous section?}

Typically managing an NFR encompasses a number of steps, including elicitation, analysis, negotiation, implementation, verification, and validation.
%However, some of these practices (e.g. do not require 
However, these practices at our organizations suggest that an NFR may be ``realized'' \emph{without direct} implementation, i.e. actions have been taken to satisfy that the conditions of the NFR have been met, although it may not necessarily be implemented or verified by the organization.
Hence we use ``realization'' as a broader term, as opposed to the traditional implementation. 
The ``realization'' of an NFR is composed of a number of sub-tasks and in this paper we use realization to indicate when an organization has reached a satisfactory level of an NFR to have ``realized'' it, whether or not it is completely `satisficed' or not.
% \colin{R1C1a}
\hghlt{However, during the process of ``realizing'' an NFR, it is vital to ``put a number on the NFR'', as a ``realized'' NFR may be affected, perhaps negatively, unbeknownst to developers (e.g. the implementation of a new feature causes \nfr{performance} to crater).}
%; as an organization can ``realize'' an NFR \emph{without} directly implementing it. 

While our study did not directly observe the elicitation of NFRs at the three organizations, the practices we identified were concrete actions these organizations took to support NFR realization and verification.
While the recent comprehensive SLR \cite{yu2020utilising} confirmed the ability to \emph{verify} NFRs by leveraging CI, this is just one aspect of NFR management.
%\omar{Again with CI...}
NFR verification may confirm \emph{if} an NFR has been realized; however, realizing an NFR doesn't necessarily mean the NFR is verified.

For example an organization may realize, potentially only a part of, \nfr{resiliency} by offloading it to Amazon Web Services and potentially verify \nfr{resiliency} with some form of chaos engineering \cite{basiri2016chaos}; however, note that realization and verification do not necessarily go hand-in-hand.
As part of NFR management, we found an organization was able to realize an NFR, for example \nfr{availability}, \nfr{resiliency}, or \nfr{scalability} by offloading to a third-party, ultimately resulting in very little overhead (aside from cost) to the organization.
%\omar{Was that intentional, or was that a side-effect of offloading?}

While the SLR by Yu et al. \cite{yu2020utilising} is the closest work to ours, they found leveraging CI is underutilized to verify NFRs and that the ratio of industrial to theoretical studies is low---thus highlighting the importance of our study bringing substantial empirical evidence to support that CSE is an enabler in, not only testing but, realizing NFRs.
Furthermore, through our study of the practices and challenges at these organizations we uncovered 30 NFRs that they found relevant (which are clearly not complete for \emph{all} organizations).
Seven NFRs are in common with the findings from Yu et al. \cite{yu2020utilising} (\nfr{latency} and \nfr{productivity} being the exceptions).
Notably absent from their list are 3 of our top-5 NFRs, namely \cm{}, \nfr{security}, and \nfr{usability}.

% \neil{consider dropping this paragraph entirely}
% \omar{To Neil: Why? I think it drives the point in.}
Although Yu et al. specifically mention \nfr{usability} as hard to verify, our study provides evidence to suggest \hll{that some organizations are satisfied with their level of realizing \nfr{usability}; of course this distinction is relative and may not apply to other organizations in such a black and white manner.}
In particular, we previously discussed how Beta was able to leverage CSE to realize \nfr{usability} in real-time  (see Section \ref{sec:find_own_tool}).
In addition, Gamma is able to track \nfr{usability} metrics through their CSE practice, including user events, such as button clicks, page views, and navigation traces, and runs large scale A/B experiments \cite{Gupta2018}.
\hll{While, this distinction certainly merits further investigation to \emph{exactly} how this organization satisfactorily realizes \nfr{usability} and how this realization can be applied to other domains and organizations, the exact details are outside the scope of this paper.}
% These metrics are used to statistically quantify, monitor, and realize \nfr{usability} in real-time.
By leveraging metrics, the feedback loop, and continuous monitoring, Beta and Gamma have near-constant realization of \nfr{usability}---an otherwise difficult to realize NFR.

% The recent SLR by Yu et al. \cite{yu2020utilising} examined how organizations are leveraging CI to test NFRs.
% \todo{This is a bit weak. How can we make this stronger?}
% Yu et al. uncovered a total of 9 NFRs, of which 6 are in common with our top-10: \nfr{availability}, \nfr{maintainability}, \nfr{performance}, \nfr{reliability}, \nfr{scalability}, and \nfr{stability}.

Metrics, of any kind, are the starting point that allow an organization to set goals, track progress, and monitor the state of the system in a reliable fashion \cite{neely_continuous_2013}.
\hll{However, metrics are not without problems, as assigning a desired threshold to an NFR is not trivial.
Fixed or static thresholds may be problematic for complex NFRs, requiring alternative solutions such as desired, minimum, dynamic thresholds \cite{rehmann2016performance}, or even the use of artificial intelligence to adapt the thresholds.}

We believe that continuous, rapid iterations using metrics, the feedback loop, and continuous monitoring brings an increase to transparency and traceability of NFRs.
% Is realizing an NFR important to them?
Transparency and traceability are afforded by allowing anyone in the organization to easily track changes to a particular NFR metric, back to a localized source commit in the code \cite{rath2018traceability}.
Traceability in CSE has been previously studied in the Eiffel approach \cite{Sthl2016}; however, the Eiffel approach is aimed at improving the CSE pipeline, not necessarily the resulting software.
The authors \cite{Sthl2016} note that further work includes extending Eiffel to consider development activity, which would ideally include NFR activities as well.

% Yu et al. concluded that while only nine NFRs were uncovered through their SLR, other NFRs are still unknown and merit future research. 
% We highlight the importance of one such NFR, \cm{}, in Section \ref{sec:cfg_mgmt}.
% In addition, while Yu et al. explicitly noted that some NFRs, such as \nfr{usability} are hard to automate and test, we provide evidence to suggest otherwise.

The ability to realize and continuously monitor, track, and audit NFRs in real-time throughout the entire life cycle of a software project is immensely powerful \cite{Sthl2016}.
Alternatively, an organization may hire consultants to assess satisfaction of a particular NFR, such as \nfr{security}; however, this is often a one-time assessment and does not help with \emph{continuous}, ongoing satisfaction of the NFR in question, which is usually the key from an operational point of view.
%\omar{The previous paragraph feels like we're beating a dead horse.}

While assigning metrics to NFRs is not a new idea and has always been a recommended practice to ensure proper verification of NFRs \cite{bass_software_2012}, consideration of the metric often only happens during the initial design and architecture phases.
Once the NFR has been defined, measured (and perhaps satisfied), and the organization is deeply entrenched in actual coding, the NFR may no longer be tracked \cite{caracciolo2014software}.
The key novelty with CSE is that it facilitates realizing NFRs and the constant and continuous ability to monitor and satisfy NFRs through the quick feedback loop.
An organization is able to look at their CSE pipeline and determine the gap between NFR objectives, and actual level of \nfr{performance}, \nfr{usability}, or \cm{} (among others).

\begin{mybox}[Research Implication 1]
%===============================================================================
%Our research has highlighted four practices that these three organizations used to realize NFRs \hll{(to varying degrees),} including otherwise difficult to verify NFRs such as \nfr{usability}, in CSE.
%\omar{Maybe elaborate a bit more on how CSE did the enabling? Just the important characteristic that we're arguing for (rapid feedback and continuous behavior)}
\hll{Our research has highlighted how CSE has enabled organizations to realize NFRs (to varying degrees) through the four practices highlighted in our research.}
% Further empirical research to expand our understanding is needed to 
% Future research to expand our understanding of other potential NFRs that can be satisfied by leveraging CSE is required.
\hll{While realizing NFRs in CSE is a promising trend, future empirical studies should seek to investigate the impact of continuously monitoring NFR satisfaction on software design.}
%expand and validate our understanding of these practices by studying more organizations.
%\neil{what about ``investigate the impact of continuoulsy monitoring NFR satisfaction on software design?''}
%\dana{Neil, i like your suggestion; just not sure how to make it flow with the first part of the implication. thoughts? }
%we need to validate if the four practices actively encourage and enable an organization to focus on satisfying NFRs.
%In addition, future research into whether our practices demonstrate the importance of treating NFRs in a similar fashion as functional requirements would be worthy.
%Due to the extremely low number of studies on NFRs in CSE there is a need for additional research to continue furthering the advancement of NFR treatment in a CSE context.
%===============================================================================
\end{mybox}
%===============================================================================

%===============================================================================
\subsection{Trade-Offs in Realizing Non-Functional Requirements in Continuous Software Engineering}
%include 5.3, 5.7, 5.8
%===============================================================================
While we have shown that CSE further enables an organization to realize NFRs, from our findings we uncovered three notable trade-offs.
For each trade off, we discuss our findings in relation to relevant existing literature and highlight areas worthy of further research.

%BENEFITS OF OFFLOADING
\subsubsection{Offloading NFRs to Third-Party Providers Results in Losing Control Over an Offloaded NFR}
%The first trade-off is the cost of losing \emph{control} over an offloaded NFR; of which all three organizations were able to successfully offload NFRs to third party services, providers, or tools.
The emergence of cloud providers, such as AWS, Google Cloud, and Microsoft Azure, offers significant advantages to software development organizations. 
First, it has encouraged and facilitated small organizations to realize NFRs, \hll{perhaps only partially through sub-tasks,} that would otherwise not be within their reach, such as \nfr{scalability}. 
Second, offloading \hll{sub-tasks of} NFRs, such as \nfr{scalability} and \nfr{performance}, enables an organization to devote additional resources to enhancing the core product \cite{anderson2019performance} and is key to a small organization's business success. 
Often, finding money to pay for NFR offloading is easier than finding staff or time, especially for small, resource constrained organizations.
Furthermore, there is some notion that some NFRs may be realized and guaranteed through certified quality of service guarantees \cite{anisetti2020cost}, allowing an organization to focus on the core of their business. 
Third, the utilization of cloud platforms \cite{nouacer2016equitas}, or even simulators \cite{soni2015end}, allow an organization to easily build otherwise costly environments solely for the purpose of verifying \nfr{reliability}, \nfr{availability}, \nfr{performance}, and \nfr{scalability}. 
%The use of cloud platforms has popularized another form of NFR verification, chaos engineering \cite{basiri2016chaos}, for example to verify \nfr{resiliency} of a system an engineer randomly terminates virtual machines and observes how the system behaves.
%Chaos engineering also helps an organization build confidence in offloading an NFR, which may help overcome the cost of offloading an NFR.

\hll{At the same time we must recognize that offloading an NFR may not imply the NFR is realized across all aspects of the organization.
This is even more important with the prevalence of distributed or micro-service architectures, as offloading an NFR for one particular component does not satisfy that NFR for the entire system.
Given the cross-cutting nature of NFRs, one must recognize the limitation of offloading an NFR and carefully plan to ensure that offloading will actually achieve a desired result.
%Furthermore, another limitation of offloading is when an NFR, such as \nfr{security} is offloaded on one or more components does not guarantee it is achieved across \emph{all} components.
%Offloading an NFR is not a simple binary black or white, in a similar to manner to realizing an NFR.
Organizations must be cognizant of the limitations of offloading an NFR.}

We identified two costs associated with offloading an NFR, 1) the loss of control of the NFR and 2) the potential for vendor lock-in. 
First, the offloading organization will be at the mercy of the organization taking over that NFR \cite{cusumano_cloud_2010}.
If an NFR is realized by decomposing that NFR into a series of sub-tasks, then an organization might lose control of those sub-tasks, or the assigned priority of those sub-tasks.
In particular, if an organization has offloaded \hll{a portion of and} NFR, such as \nfr{availability}, to a cloud provider and that cloud provider experiences an issue, such as an outage, the organization will also experience an outage and hence the \nfrno{availability} of the organization's product is now entirely out of their control \cite{benlian_opportunities_2011}.

%By relying on a third party tool or service to manage \hll{a portion of an} NFR, such as \nfr{security}, the additional layer of abstraction may reduce the familiarity and understanding of how the third party treats an NFR, e.g. \q{Identity and access management [...] that's within Amazon because we have far too many things to keep track of} (P3).
%Similarly, if a cloud provider has a security vulnerability, then the security issue will likely propagate to any organization using that third party, and there is very little the organization can do to rectify the issue, except wait.
%To the best of our knowledge, the loss of control of an NFR has not yet been studied and is worthy of future work.

% ANOTHER TRADEOFF OF OFFLOADING IS THE POTENTIAL FOR VENDOR LOCKIN
Second, with offloading there is also the risk of vendor lock-in, which occurs when a customer is overly dependent on a vendor, such as a cloud provider, and is unable to switch to another vendor without substantial re-work. 
Vendor lock-in has long been a problem in the software industry \cite{mainframe95}; interestingly, we did not hear about vendor lock-in from any of our interviewees.
% However, neither cloud provider tools nor standards are widely adopted (see Section \ref{sec:cfg_mgmt}) so the potential for vendor lock-in remains, and is a substantial research gap \cite{guerriero2019adoption}.
\hll{However, neither cloud provider tools nor standards are widely adopted (see Section \ref{sec:cfg_mgmt}) so the potential for vendor lock-in remains and is an area of active research \cite{guerriero2019adoption, opara-martins_critical_2014, opara-martins_critical_2016}.}

% realizING AN NFR IN CSE MAY RESULT IN LSU OF SAID NFR
% THE IMPORTANCE OF SHARED UNDERSTANDING AND HOW OUR PRACTICES FACILITATE IT
\subsubsection{CSE Hurts Shared Understanding of the NFR}
% 1. mostly the same
% 2. we believe there might be a LSU
% 3. talk about the in-depth
% 4. talk about explicit/implicit how it corroborates Glinz
% 5. then what do we do with 3/4!
While shared understanding is a critical success factor in producing high quality software designed to meet stakeholders' needs \cite{bittner_why_2013}, our study highlighted a lack of shared understanding of key NFRs in each of the three organizations, despite their ability to leverage CSE to realize the NFRs. 
For example, our organizations described how role siloing between DevOps and developers creates a lack of shared understanding of NFRs.
However, this phenomenon lacks substantial empirical research \cite{glinz_shared_2015}, as the practice of creating and maintaining a shared understanding in agile is not well established \cite{schon_agile_2017}.
%Thus, there is merit for a deeper study on the trade-off of realizing an NFR by leveraging CSE and the effect on the shared understanding of that NFR.

In a follow-up investigation on this trade-off, we sought to further understand and quantitatively validate details of this lack of shared understanding from an analysis of the project management repositories at these organizations \cite{werner2020re}. 
%Drawing on previous literature that indicates that lack of shared understanding leads to rework \cite{damiantse2006, bjarnason2011case}, we conducted an in-depth analysis of 41 NFR-related rework tasks across the three organizations. 
%These 41 one tasks were selected from a large set of tasks and confirmed, through a series of focus groups, to be rework caused by a lack of shared understanding of an NFR.
%We performed a further empirical investigation on this trade-off, through the in-depth analysis of 41 NFR-related rework tasks at these three organizations practicing CSE \cite{werner2020re}.
%To identify the lack of shared understanding we took advantage of how the lack of shared understanding has been shown to result in substantial rework \cite{damiantse2006, bjarnason2011case} and represents 40-50\% of the effort on software projects \cite{boehm2005software}.
%Through focused user groups, each of the 41 tasks were confirmed to be rework caused by a lack of shared understanding of an NFR. '
%\omar{The previous and next paragraphs; why is here the first time I'm reading about the task management analysis? Shouldn't we mention that in the methodology?}

Our analysis identified that while there is an acceptable and unavoidable amount of lack of shared understanding, which captures unknown unknowns \cite{sutcliffe-unknowns} and represents desirable learning and feedback \cite{riesStartup}, 78\% of the lack of shared understanding was deemed \emph{avoidable} by the three organizations.
These results bring additional evidence that an organization realizing NFRs in CSE may do so at the cost of a lower shared understanding of those NFRs.

\subsubsection{Fast Pace of CSE Deprioritizes NFRs}
Agile methodologies have been shown to risk the overemphasis of FRs at the expense of NFRs \cite{ramesh_agile_2010}.
Our findings corroborate previous research \cite{savor_continuous_2016, gralha_evolution_2018} indicating that the fast pace of CSE increases the risk of deprioritizing NFRs.
Our results also suggest that neither frameworks nor models produced from research are adopted in practice to assist prioritizing NFRs.

% THERE IS A LACK OF PRIORITIZATION OF NFRS (in part due to LSU)
%Research has produced numerous frameworks and models developed specifically for NFRs, including elicitation \cite{maalej2015toward, groen2017crowd, zowghi2005requirements}, specification \cite{caracciolo2014software, behutiye2017non}, traceability \cite{furtado2016trace, cleland2005goal}, and prioritization \cite{paucar2017arrow, misaghian2018approach}.
%Albeit we found that no such framework or model was used by any of the three organizations.

%The underlying issue is a lack of domain or technical knowledge of NFRs to even acknowledge the necessity of such a framework or model.
While the developers we interviewed indicated that NFRs are important (to developers), the perceived importance of NFRs differs for product managers, among others.
As such, NFRs were largely left to the developers to self-manage in an ad-hoc manner.
%The priority of NFRs at these three organizations is so low, as NFRs are underspecified and largely neglected---until of course they become a critical issue.
Despite the numerous frameworks and models developed through research \cite{maalej2015toward, groen2017crowd,behutiye2017non, misaghian2018approach}, there exists a gap between industry and practice on whether they can actually be used to solve this issue of NFR prioritization.
This is a significant empirical finding adding to the scarce evidence on how (the lack of) NFR prioritization is handled in industrial versus research settings.
\begin{mybox}[Research Implication 2]
%===============================================================================
% Competitive software market forces push organizations to adopt the practice of offloading NFRs to third parties, despite the loss of control over those NFRs. 
% An organization's viability may hinge on the convenience of NFR offloading offered by third party providers. 
%realizing an NFR through CSE comes at the cost of substantial trade-offs, such as loss of control, lack of shared understanding, and lower priority of NFRs.
Realizing an NFR through CSE comes at the cost of substantial trade-offs, such as lower priority and a lack of shared understanding of NFRs, or for organizations leveraging offloading to third-parties, loss of control of the NFR. 
Research is needed to develop and empirically evaluate mitigation strategies or techniques to reduce the risk of and to overcome these trade-offs.
%===============================================================================
\end{mybox}
%===============================================================================

%===============================================================================
\subsection{The Importance of \cm{} as a Non-Functional Requirement}
\label{sec:cfg_mgmt}
%include 5.5
%===============================================================================
% DEFINE CONFIGURATION MANAGEMENT and what it entails
% When an organization adopts CSE practices, an NFR that rises in importance is configuration management. 
% \cm{} facilitates the management and control of an organization's software infrastructure  \cite{bass_software_2012} often through scripting in the form of ``Infrastructure as Code''.
% In particular, \cm{} includes configuring organizational build, staging, and deployment infrastructure, ideally with little to no human intervention \cite{humble2018accelerate}.
% \cm{} is a vital NFR and is a key contributor to the perceived benefits of CSE \cite{humble2010continuous}.
% Yet, \cm{} as an NFR receives insu 

The importance of \cm{} as an NFR grows as an organization relies more heavily on CSE practices.
\cm{} is an attribute of the software system that refers to how easily an organization can configure its software infrastructure and environments \cite{bass_software_2012}, including ``Infrastructure as Code'' (IaC).
% \colin{address R1C1b}
\hghlt{However, \cm{} is \emph{more} than just a system quality, as it also encompasses overarching \emph{process quality.}}
Our data show the importance of the non-functional quality of the system's configurability---the source code, build scripts, infrastructure and deployment configuration, and associated hardware. 
Like maintainability, configurability refers to an internal quality that supports the goal of rapid deployment and re-configuration. 
To realize this NFR, one might use tactics such as rollbacks, keeping production and development environments in sync, or applying infrastructure as code tools such as Puppet.

Comprehensive \cm{} has long been considered to be an enabler of the many perceived benefits of CSE by Humble et al. \cite{humble2010continuous}.
However, \cm{} is largely underrepresented.
%\textcolor{red}{it is not considered an NFR in the recent SLR by Yu et al. on continuous NFR testing \cite{yu2020utilising}.}
The concept of \cm{} has further grown to encompass the configuration of build, staging, and deployment infrastructure, ideally with little to no human intervention \cite{humble2018accelerate}.
% ORGANIZATIONS NEED TO SPEND ADDITIONAL EFFORT ON CONFIGURATION MANAGEMENT -- lack of accepted standards.
%While \cm{} is usually stored in source control, \cm{} has not traditionally received the same attention as traditional application code; however, Humble et al. indicate that this is a misconception \cite{humble2018accelerate}.
Our study brings clear evidence that \cm{} should be considered an extremely important and high priority NFR in CSE.
%is the most mentioned NFR \histno{ConfigurationManagementNFR}{103} and, hence, 

% EXAMPLES OF HOW CONFIGURATION MANAGEMENT IS AN ENABLER
As software systems increasingly exist as a service running in the cloud, application code is no longer the only important source code. 
Infrastructure configuration and code are as vital to software business goals as application code. 

At Alpha, customers are in part paying for Alpha to host reports and data for them.
Thus, their infrastructure configuration and code must also exhibit NFRs such as \nfr{reliability} and \nfr{availability}. 
In contrast, these NFRs are entirely the customer's responsibility for an on-premise offering.
An organization must now invest in \cm{} in parallel with other NFRs and features.

%\omar{Does specifying the job title at a particular org compromise that person's anonymity? Do the orgs know which Greek letter they've been assigned?}
As the technology director at Gamma commented during our member checking phase, \q{[\cm{}], and associated IaC and automation is the enabler that allows organizations like ours to essentially offload other NFRs such as \nfrno{availability}, \nfrno{scalability}, and \nfrno{security} to cloud providers [...] without [\cm{}] other NFRs would suffer, such as \nfrno{reliability}, \nfrno{maintainability}, \nfrno{repeatability}, and even \nfrno{availability} due to more human error during deployments.}

% HOW WE MUST TREAT CONFIGURATION MANAGEMENT AS A TOP PRIORITY NFR susceptible to all the already-documented issues with NFRs
The increased reliance on \cm{} results in a trade-off: the organization needs to now spend significant additional resources on configuration, developer training, and avoiding potential vulnerabilities associated with \cm{}, including the lack of shared understanding.
%It is important for an organization to realize the large onus on an organization due to the importance of \cm{} as NFR and treat \cm{} as important as application source code, as it is susceptible to the same errors or vulnerabilities \cite{jiang2015}.

First, \cm{} now has its own set of distinct code-smells \cite{schwarz2018code}.
There are early efforts to mitigate these code-smells, such as Rahman et al. \cite{Rahman2019} to identify code-smells of \cm{} code in open source software.
Second, while \cm{} may benefit from standard coding practices it is not yet done in practice \cite{guerriero2019adoption}.
Conversely, \cm{} is actually associated with a wide variety of disparate languages and tools.
Most organizations use three or more different tools and no single tool is used by the majority of organizations \cite{guerriero2019adoption}.
%\neil{consider removing the standards stuff ...}
%\omar{This next bit feels disjointed.}
Existing standards, such as Topology and Orchestation Specification for Cloud Applications (TOSCA) and Open Cloud Computing Interface have been proposed; although the adoption amongst DevOps engineers is low (18\%) \cite{guerriero2019adoption}.
Third, the standards themselves do not contain a complete set of NFRs, as they require extensions to include \nfr{security} \cite{saatkamp2017topology, shu2017study} and \nfr{privacy} \cite{7973748}, among others.

Our study highlights the need to fill in the gap between research and industry efforts on supporting \cm{}.
As more and more configuration is stored as code (IaC), the same problems we see in traditional NFRs is likely to surface in \cm{}, including the aforementioned trade-offs (e.g. loss of control and shared understanding). 

%Concepts such \nfr{portability}, \nfr{usability}, \nfr{extensibility}, \nfr{learnability}, and \nfr{stability} must now be considered alongside \cm{} \cite{guerriero2019adoption}.
% MOTIVATING THE EFFORT DUE TO THE IMPORTANCE OF CONFIGURATION MANAGEMENT

%===============================================================================
\begin{mybox}[Research Implication 3]
%===============================================================================
%\cm{} emerged as a prominent NFR for the organizations in our study. 
\cm{} is emerging as a vital NFR and an integral part of developing software; however, it is yet understudied as an NFR. 
%For example, it is not mentioned in the recent SLR from Yu et al. \cite{yu2020utilising}.
%Given the importance of \cm{} in CSE,
In-depth research is required to develop and evaluate techniques to manage \cm{}, including elicitation, analysis, validation, and verification.
Further empirical studies are needed to explore and propose solutions to the associated trade-offs and challenges with \cm{}.
%We previously highlighted the importance of overcoming trade-offs and challenges of NFRs in CSE, in general, and with the revelation of the importance of \cm{} as an NFR empirical research is needed to explore techniques to overcome trade-offs and challenges associated with \cm{}.
%However, we do not yet know the extent of \cm{}'s importance for larger organizations nor other industrial contexts.
%As more organizations rely on CSE practices, organizations require methods to the associated trade-offs of \cm{} as an NFR.
% consider ditching this
% Researchers may find it prudent to investigate how widespread the use of configuration management is and the impact of \cm{}, including common faults and exactly which NFRs may be realized through configuration management.
% \todo{Strengthen the argument that ConfigMgmt itself is an NFR of utmost importance in CSE}
% 2nd theme --> tradeoffs; CSE increases LSU; so orgs should mitigate LSU by NFRs. Now since config mgmt is an NFR, we need to explore techniques to enhance LSU, etc (related back to the first two themes.)
%===============================================================================
\end{mybox}
%===============================================================================

\subsection{Implications for Practitioners}
In addition to researchers, our study has wide reaching implications for practitioners. 
Our observed practices and challenges for handling NFRs in CSE demonstrate that organizations are both successfully treating NFRs and encountering difficulties.
For practitioners, there are three noteworthy implications. 
First, organizations must be aware of the ability to realize NFRs using the four practices that leverage CSE, but also be mindful of the associated challenges.
Second, while offloading NFRs to a third party provider has many potential benefits, practitioners should be aware of the potential consequences of offloading. 
Furthermore, practitioners should monitor any offloaded NFR to ensure the NFR is treated as expected.
Finally, practitioners need to recognize the importance of \cm{} and dedicate time to educate, elicit, analyze, and verify \cm{}.
%\omar{This subsection feels a little light. What other claims (or elaboration on existing claims) can we make here based on our findings, and why should practitioners care about this?}

% \subsection{TODO}
% \todo{IGNORE THIS}
% In case we need more...

% Some of the NFRs are operationalized in domain specific technologies, which may inhibit the generalizations.

% Ideally, an organization has a fast feedback loop so that the organization can quickly pinpoint problematic NFRs or alter an NFR based on customer response.

% read \cite{pietrantuono2019towards}
% READ \cite{ferme2018declarative}

% \cite{knauss2016continuous} there are already a plethora of tools in CI and adding additional tools to help NFR verification only further destabilize the CI pipeline

% \cite{waller2015including} indicate that verification can take too long and the feedback loop is no longer quick.

% OLD SUBSECTIONS
%===============================================================================
%\subsection{Why Offloading NFRs is Important to Small Organizations}
%===============================================================================

%===============================================================================
%\subsection{How Continuous Monitoring Supports NFRs}
%===============================================================================

%===============================================================================
%\subsection{Growing Importance of Configuration Management as an NFR}
%===============================================================================

\section{Threats to Validity} 
Threats to validity in qualitative research typically concern the reliability of the results. 
We use the total quality framework of Roller \cite{Roller2015} to discuss these potential threats and our mitigation strategies. 
The framework is a way to assess the quality of qualitative research using four categories: credibility, analyzability, transparency, and usefulness.

Concerning the \emph{credibility} of our study and data gathering methods, our study investigated the state-of-the-practice at three industrial organizations performing some form of CSE. 
Our selection of these three organizations might suffer from sampling bias, as we chose organizations willing to participate from a larger pool of local companies.
However, our preparatory study phase ensured that they practiced CSE inline with industry and literature best practices to the best of our knowledge.
%\todo{I thought we dropped this next sentence?}Furthermore, we make no claims on generalizability, although we have included as many pertinent details of our methodology such that the applicability of our results may be further studied and replicated.

% I make this a comment because it is too defensive, not necessary I would say. We consider these organizations fairly representative (in our experience) of software organizations. 
%In addition, we see no reason why our results should not apply to \emph{any} organization practicing CSE, although this claim should be subsequently validated.
Our interviewees were representative of their organizations with respect to role, gender, and experience.
The unbalanced distribution of roles (12 developers; 6 managers) and genders (13 males; 5 females) is representative of our organization's demographics, and unfortunately there were no other managers or females to interview.
Our analysis did not reveal differences due to gender, role, or experience; this represents a worthwhile direction for future study. 
%We acknowledge that females are underrepresented (13 males; 5 females); however, we were constrained by the respective organization's demographics and their respective availability.
%Finally, while we recorded role, gender, and experience for each interviewee, no comparison was made between the various groups, although this could may be useful for future work.
%We interviewed 18 of the approximately 150 employees at these organizations, and reached saturation (lack of novel insights) at each organization, so adding more participants likely would not have changed our findings. 

To mitigate the threat of construct validity, we began each interview by examining the respondent's knowledge of NFRs. 
We then explained the NFR concept with examples so that each respondent had a similar level of understanding about NFRs.
All our participants were proactive and valuable in offering details commensurate to their role and experience with their organization's practices.

As far as the credibility associated with data analysis, the primary threat is in the coding approach. We described our process in Section \ref{method}. 
We followed best practices for thematic coding and used the inter-rater agreement process frequently to align coding schemes.
\hll{Due to the number of coders we allowed multiple codes to be applied to a unit, thus limiting our ability to apply an inter-rater agreement that would resolve chance agreement.}
Finally, we may be susceptible to researcher-participant interactions, since these were in-person interviews. 

As for \emph{analyzability}, we used computer-aided transcription, but we did check each transcript against the audio where the transcription was unclear. 
We utilized the open coding approach to remove the potential bias from coders.
We also performed peer debriefings and analyzed deviant codes to verify our analysis and ensure our results were consistent and neutral.

With respect to \textit{transparency}, we used histograms to enhance our thick descriptions of the responses. 
For reliability, we also conducted a member checking exercise to validate our findings with our subjects (12 of 18 interviewees responded). 
We elicited ordinal feedback (Strongly Disagree-Strongly Agree) on each of our practices and challenges. 
For all 4 practices and 3 challenges, the 12 respondents had a median score of ``Agree''. 
One challenge that we had originally included had a median score of Neutral, with 5/12 voting ``Disagree'' or ``Strongly Disagree''. 
As a result, we dropped this particular challenge pending further investigation. 
We were able to integrate additional insight (from one Director of Technology at Gamma) into our discussion of findings. 
We also make our codebook and analysis scripts available for replication in our research artifacts repository
but due to NDA, we cannot share raw transcripts. 

The \textit{usefulness} of our study is geared towards bridging the gap between research and practice of handling NFRs for CSE organizations.
In particular, we raise awareness of areas for researchers to focus on with respect to the current and emerging trends that can enable organizations to realize NFRs.
\hll{We recognize that NFRs cannot and should not be grouped together, as the differences between individual NFRs can be as great, if not greater, than the difference between a FR and NFR; we believe that further in-depth studies should be focused on individual NFRs.}
While the usefulness to practitioners is to help bring focus to \emph{how} an NFR can be realized and the associated pitfalls to each.
\section{Conclusion}
The effective management of NFRs is key to successful, high-quality software projects.
NFRs themselves are well-known to be difficult to express, let alone manage, in part due to their cross-cutting nature.
Since NFRs in the context of CSE have not been sufficiently explored in literature, we conducted a qualitative study to gather empirical evidence on how CSE organizations handle NFRs.
%\hll{We conducted an exploratory study to elicit hypothesis from practice, which we aim to validate in future work.}
Contrary to previous research, our investigation brings insights from three organizations that do manage NFRs using a variety of practices, yet continue to face important challenges and make trade-offs.

By discussing the practices and challenges from our findings, we also formulated research implications both for research and practitioners. While NFRs are difficult, ambiguous, and tough to verify in normal circumstances, we believe following the four practices will allow an organization to realize NFRs in CSE.
In particular, our empirical evidence indicates that a key to rein in NFRs is to leverage CSE practices, such as the quick feedback loop or the capability to offload NFRs to third-parties. However, the peril of realizing an NFR by leveraging CSE is that an organization may lose control of an offloaded NFR, leaving them at the mercy of the third-party, or incur a decrease in the shared understanding of an NFR.

The practices and research implications we presented in this paper represent useful avenues for future research, in the form of hypotheses or methods to be validated through empirical studies in a broader set of, potentially larger, CSE organizations.
%While our research studied smaller CSE organizations, we need to examine whether implications are applicable to a broader set of, potentially larger, CSE organizations; however, some of the practices (such as offloading an NFR) may have more profound effects on a smaller, resource constrained organization.
%\neil{weak ending}
%\omar{I agree with Neil on this. As Peggy and Neil keep telling me, we can be a bit brave with respect to the claims we're making here.}
%Some NFRs are more difficult to operationalize than others; as researchers we need to systematically identify, for each NFR, how to operationalize that NFR in CSE and, more importantly, how to potentially generalize the approach to other NFRs.
%We've reported on a number of NFRs, some of which are difficult to operationalize.
%However this list of NFRs is by no means exhaustive as there are many other NFRs that merit future research.
%We need to explore what industry is doing and how these techniques can be generalized to other organizations.
%By doing so we will be able to identify the gap containing NFRs that have no known solution.

%\begin{acks}
\section*{Acknowledgments}
%We would like to thank our three partner companies and employees for their time and collaboration.
We thank our three partner organizations for their time and collaboration.
Our research was supported by Canadian grant NSERC-CRD 535876.
%\end{acks}
%\newpage
%\bibliographystyle{ACM-Reference-Format}
\bibliographystyle{IEEEtran}
\bibliography{main}

%\listoftodos{}

\end{document}